\documentstyle[aps,prb,floats,amsmath]{revtex}

\def\E{{\cal E}}
\def\eps{\epsilon}
\def\half{{\textstyle{1\over2}}}
\def\chibar{\bar{\chi}}
\def\Cbar{\bar{C}}
\def\Sbar{\bar{S}}
\def\ebar{\bar{e}}
\begin{document}
\draft

\twocolumn[\hsize\textwidth\columnwidth\hsize\csname
@twocolumnfalse\endcsname

\title{ Systematic treatment of displacements, strains and
electric fields in density-functional perturbation theory}

\author{Xifan Wu$^{1}$, David Vanderbilt$^{1}$, D.R.~Hamann$^{2}$ }

\address{$^{1}$Department of Physics and Astronomy, Rutgers University,
Piscataway, New Jersey 08854-8019, USA}

\address{$^{2}$Bell Laboratories, Lucent Technologies,
Murray Hill, NJ 08974}
\date{January 28, 2005}

\maketitle

\begin{abstract}
The methods of density-functional perturbation theory may be used
to calculate various physical response properties of insulating
crystals including elastic, dielectric, Born charge, and
piezoelectric tensors.  These and other important tensors may be
defined as second derivatives of the total energy with respect to
atomic-displacement, electric-field, or strain perturbations, or as
mixed derivatives with respect to two of these perturbations.  The
resulting tensor quantities tend to be coupled in complex ways in
polar crystals, giving rise to a variety of variant definitions.
For example, it is generally necessary to distinguish between
elastic tensors defined under different electrostatic boundary
conditions, and between dielectric tensors defined under different
elastic boundary conditions.  Here, we describe an approach for
computing all of these various response tensors in a unified and
systematic fashion.  Applications are presented for two materials,
wurtzite ZnO and rhombohedral BaTiO$_{3}$, at zero temperature.
\end{abstract}


\vskip2pc]

\narrowtext
% \marginparwidth 2.9in
% \marginparsep 0.5in
% \def\dvm#1{\marginpar{\small DV: #1}}
% \def\xwm#1{\marginpar{\small XW: #1}}
% \def\dhm#1{\marginpar{\small DH: #1}}

%%%%%%%%%%%%%%%%%%%%%%%%%%%%%%%%%%%%%%%%%%%%%%%%%%%%%%%%%%%%
\section{INTRODUCTION}
%%%%%%%%%%%%%%%%%%%%%%%%%%%%%%%%%%%%%%%%%%%%%%%%%%%%%%%%%%%%

The methods of density-functional theory (DFT)\cite{KS} and
density-functional perturbation theory (DFPT)
\cite{Gonze,Baroni}
have been shown to give a successful description of the dielectric
and piezoelectric properties of a wide range of materials in which
electronic correlations are not too strong.\cite{Dielectric,Piezoelectric}
Many properties of interest can be computed directly from DFT using
finite differences -- for example, elastic constants computed from the
stress arising from a small applied strain, or dynamical
effective charges computed from polarizations\cite{ksv}
arising from small sublattice displacements.  On the other hand,
the use of DFPT methods is becoming increasingly popular because
it can be used to compute such response properties directly,
without the need for multiple ground-state calculations, thus
providing the desired response properties in a more automated,
systematic, and reliable fashion.

As a result, improved DFPT capabilities have been implemented in
recent years in several of the computer code packages commonly used
by the computational electronic-structure
community.\cite{ABINIT,PWSCF,VASP} This development has been most
thorough in the case of the open-source ABINIT computer
package,\cite{ABINIT} in which the capability for handling strain
perturbations \cite{Hamann} has recently been added to the previous
implementation of atomic-displacement and electric-field
perturbations.  This development opens the prospect for a systematic
treatment of three kinds of perturbations in insulating crystals
on an equal footing: periodicity-preserving atomic displacements
(i.e., zone-center phonons), homogeneous electric fields,
and homogeneous strains.  These three degrees of freedom are often
strongly coupled, especially in polar materials used in
modern ferroelectric, piezoelectric, and dielectric applications.

In this work, we show that such a systematic approach is now not
only practical, but especially powerful.  That is, we show that
computing the full set of six elementary (or ``bare'') response
tensors (force-constant, dielectric, elastic-constant, Born-charge,
internal-strain, and piezoelectric tensors) associated with these
three kinds of perturbations provides an extremely valuable
database that can subsequently be used to compute a wide variety of
relevant physical properties.  Among these, for example, are the
physical dielectric, elastic, and piezoelectric tensors (in which
atomic displacements have been taken into account), elastic
compliances, free-stress dielectric tensors,
fixed-electric-displacement elastic tensors, alternative
piezoelectric tensors, and electromechanical coupling constants.
The ability to access this wide range of secondary properties
becomes possible only after the full set elementary response
tensors has been systematically computed.

The rest of this paper is organized as follows.  In Sec.~II,
we present the formal development, defining the various
elementary response tensors and showing how other response
tensors of interest can be derived from these.
Then in Secs.~III-IV we apply our approach to
wurtzite ZnO and rhombohedral BaTiO$_3$ as two paradigmatic
example systems.  We first briefly describe the practical details of
the calculations in Sec.~III, and then present the results for
the ground-state properties, elementary response properties, and
derived response properties, in Sec.~IV.  We conclude with a summary
in Sec.~V.  A careful formulation of the theory for the case in
which strains and electric fields are simultaneously present is
deferred to the Appendix.

%%%%%%%%%%%%%%%%%%%%%%%%%%%%%%%%%%%%%%%%%%%%%%%%%%%%%%%%%%%%
\section{FORMALISM}
\label{sec:form}
%%%%%%%%%%%%%%%%%%%%%%%%%%%%%%%%%%%%%%%%%%%%%%%%%%%%%%%%%%%%

%-----------------------------------------------------------
\subsection{Elementary response tensors}
\label{sec:elementary}
%-----------------------------------------------------------

Consider an insulating crystal with $N$ atoms per unit cell.
We choose a reference state in which the lattice vectors are
${\bf a}_1$, ${\bf a}_2$, and ${\bf a}_3$, the cell volume is
$\Omega_0$, and the atomic coordinates
are $R_m^{(0)}$. Here $m$ is a composite label (atom and displacement
direction) running over $1,...,3N$, and we assume that this structure
is the equilibrium one in vanishing macroscopic electric field.

We consider three kinds of perturbations applied to such a
crystal:  (i) displacements $u_m$ of the atoms away from their
equilibrium positions, (ii) homogeneous strains $\eta_j$ where
$j=\{1...6\}$ in Voigt notation, and (iii) homogeneous electric
fields $\E_\alpha$ where $\alpha=\{x,y,z\}$ are Cartesian
directions.  We restrict our discussion to atomic displacements
that preserve the primitive-cell periodicity, i.e., to degrees of
freedom corresponding to zone-center phonon modes only.  Also, we
will restrict ourselves entirely to zero-temperature properties.

The corresponding responses that are conjugate to these three
perturbations are (i) forces $F_m$, (ii) stresses $\sigma_j$, and
(iii) polarizations $P_\alpha$.  From these, one can construct the
response functions of primary interest: ``diagonal'' responses
$K_{mn}=dF_m/du_n$ (force-constant matrix),
$\chi_{\alpha\beta}=dP_\alpha/d\E_\beta$ (dielectric susceptibility), and
$C_{jk}=d\sigma_j/d\eta_k$ (elastic constants),
and ``off-diagonal'' response tensors
$Z_{m\alpha}=dP_\alpha/du_m$ (Born effective charge),
$\Lambda_{mj}=dF_m/d\eta_j$ (internal strain), and
$e_{\alpha j}=dP_\alpha/d\eta_j$ (piezoelectric response).
However, in order to define these quantities carefully, it is
important to clarify the constraints or boundary conditions that
apply to each definition.  For example, the elastic constants
$C_{jk}$ may be defined allowing or not allowing internal atomic
displacements (``relaxed-ion'' or ``frozen-ion''), or under
conditions of fixed electric ($\E$) or displacement ($\bf D$)
field.

We take the approach here of systematically defining all response
properties as appropriate second derivatives of the energy $E$ per
unit volume with respect to the perturbations.  To be more
precise, in the presence of strains we define $E$ as the energy
per {\it undeformed} unit cell volume $\Omega_0$, while in the presence of
electric fields $E$ is modified to become an {\it electric enthalpy}
\cite{NunesGonze} by adding a term proportional to $-{\bf P}\cdot\E$,
where $\bf P$ is the electric polarization.\cite{ksv}
(While a direct treatment of finite $\E$-fields is now possible,
\cite{Iniguez,Umari} only infinitesimal $\E$-fields need to
be considered here.)  In general, we define $E$ as
\begin{equation}
E(u,\E,\eta) =\frac{1}{\Omega_0}
\left[ E^{(0)}_{\rm cell}-\Omega\,\E\cdot{\bf P}\right]
\;,
\label{eq:Etotimprop}
\end{equation}
where $E^{(0)}_{\rm cell}$ is the usual zero-field Kohn-Sham energy per
cell\cite{explan-enthalpy} of the occupied Bloch functions and
$\Omega$ is the deformed cell volume.  However, when strains
and electric fields are simultaneously present, care is needed
in the interpretation of Eq.~(\ref{eq:Etotimprop}); this is explained
in the Appendix, where a more precise formulation is given in the
form of Eq.~(\ref{eq:Etotfull}), which supersedes Eq.~(\ref{eq:Etotimprop}).
In short, the difficulty is connected with
the distinction between ``proper'' and
``improper'' piezoelectric constants;\cite{proper} we should like
our formulation to lead to the former and not the latter.  The factor
of $\Omega/\Omega_0$ has been inserted in the last term of
Eq.~(\ref{eq:Etotimprop})
towards this purpose, but this is not sufficient by itself.  In
addition, Eq.~(\ref{eq:Etotimprop}) should be rewritten in terms of
``natural variables'' $u$, $\E'$, and $\eta$, where $\E$ has been
replaced  by a {\it reduced} electric field $\E'$ that is defined in
Eq.~(\ref{eq:Eprime}).  When partial derivatives are taken with
respect to these natural variables, one automatically obtains
the ``proper'' piezoelectric tensors.  Indeed, as explained
in the Appendix, all appearances of $\E$ should
be replaced by $\E'$, with a similar replacement for polarizations,
in the remainder of this paper.  However, for the sake of clarity
of presentation, this notation has been suppressed in the main body
of the paper.

Accordingly, we provisionally write $E=E(u,\E,\eta)$ as a function
of arguments $u_m$, $\E_\alpha$, and $\eta_j$, 
with the understanding that the notation of
the Appendix supersedes the notation used here whenever strains and
electric fields are simultaneously present.  We then
expand around a zero-field reference system as
\begin{eqnarray}
E= E_0 &+& A_m\,u_m+A_\alpha\,\E_\alpha+A_j\,\eta_j
\nonumber\\
&+& \half B_{mn}\,u_m\,u_n +\half B_{\alpha\beta}\, \E_\alpha\,\E_\beta
            +\half B_{jk}\, \eta_j\, \eta_k
\nonumber\\
&+& B_{m\alpha}\,u_m\,\E_\alpha
+ B_{mj}\,u_m\,\eta_j
+ B_{\alpha j}\,\E_\alpha\,\eta_j
\nonumber\\
&+& \hbox{terms of third and higher order}
\;.
\label{eq:Eexpan}
\end{eqnarray}
We use an implied-sum notation throughout.
In this expansion, the first-order coefficients
$A_m$, $A_\alpha$, and $A_\j$ encode the
forces ($F_m=-\Omega_0\,A_m$), polarizations ($P_\alpha=-A_\alpha$),
and stresses ($\sigma_j=A_j$), respectively.
(Henceforth we shall assume that the atomic coordinates and strains are
fully relaxed in the reference system, so that $A_m=A_j=0$.)
The diagonal-block second-order coefficients
$B_{mn}$, $B_{\alpha\beta}$ and $B_{jk}$
and off-diagonal second-order coefficients
$B_{m\alpha}$, $B_{mj}$, and $B_{\alpha j}$ correspond to the
force-constant, elastic-constant, and susceptibility tensors,
and to the Born-charge, internal-displacement, and piezoelectric
tensors, respectively.

Inserting appropriate signs and cell-volume factors, the
elementary second-derivative response-function tensors are defined
as follows.  The force-constant matrix
\begin{equation}
K_{mn} = \Omega_0 \, \frac{\partial^2E}{\partial u_m \partial u_n} \, \Bigg\vert_{\E,\eta}
\;,
\label{eq:Kdef}
\end{equation}
the frozen-ion dielectric susceptibility
\begin{equation}
\chibar_{\alpha\beta} = - \, \frac{\partial^2E}{\partial\E_\alpha \partial\E_\beta} \, \Bigg\vert_{u,\eta}
\;,
\label{eq:chibardef}
\end{equation}
and the frozen-ion elastic tensor
\begin{equation}
\Cbar_{jk} = \frac{\partial^2E}{\partial\eta_j \partial\eta_k}\, \Bigg\vert_{u,\E}
\label{eq:Cbardef}
\end{equation}
are the elementary diagonal-block tensors, while the off-diagonal blocks are
the Born dynamical effective charge tensor
\begin{equation}
Z_{ma} = -\Omega_0 \, \frac{\partial^2E}{\partial u_m \partial\E_\alpha}\, \Bigg\vert_{\eta}
\;,
\label{eq:Zdef}
\end{equation}
the force-response internal-strain tensor
\begin{equation}
\Lambda_{mj} = -\Omega_0 \, \frac{\partial^2E}{\partial u_m \partial\eta_j}\, \Bigg\vert_{\E}
\label{eq:Lamdef}
\;,
\end{equation}
and the frozen-ion piezoelectric tensor
\begin{equation}
\ebar_{\alpha j} = - \, \frac{\partial^2E}{\partial\E_\alpha \partial\eta_j}\, \Bigg\vert_{u}
\; .
\label{eq:ebardef}
\end{equation}

The bar on quantities $\chibar_{\alpha\beta}$, $\Cbar_{jk}$,
and $\ebar_{\alpha j}$ indicates a frozen-ion quantity, i.e.,
the fact that atomic coordinates are not allowed to relax as the
electric field or homogeneous strain is applied.
Note that the clamped-ion elastic tensor $\Cbar_{jk}$ and
piezoelectric tensor $\ebar_{\alpha j}$ are generally not physically relevant
quantities, except in cases of high symmetry where atomic displacements
do not occur to first order in strain.  The clamped-ion susceptibility
tensor $\chibar_{\alpha\beta}$ is the purely electronic one that is measured
in response to AC or optical fields at frequencies well above the
phonon-frequency range (corresponding to $\epsilon^\infty$ in the polariton
language).

The force-response internal-strain tensor $\Lambda_{mj}$
must be distinguished from the displacement-response internal-strain tensor
$\Gamma_{nj}=\Lambda_{mj}(K^{-1})_{mn}$ that describes the first-order
displacements resulting from a first-order strain; both occur in the
literature, frequently without careful differentiation.
The piezoelectric tensor $e_{\alpha j}$ (often denoted
alternatively as $c_{\alpha j}$) describes the change of polarization
arising from a strain, or a stress arising from a change of
$\E$-field, while the $d$, $g$, and $h$ piezoelectric tensors are
defined under different constraints and have slightly different
physical meanings.\cite{Ballato}  Finally, we remind the reader that
there is some subtlety in the definition of the piezoelectric tensors
related to the specification of the energy functional when both fields
and strains are present, leading to a distinction between ``proper''
and ``improper'' piezoelectric constants\cite{proper} as will be
discussed more fully in the Appendix.  Throughout this paper, we adopt
the convention that all piezoelectric tensors are ``proper'' ones
unless otherwise noted.

We shall refer to the quantities defined in
Eqs.~(\ref{eq:Kdef}-\ref{eq:ebardef}) as the ``elementary'' or
``bare'' response tensors.  These are the quantities that will be
calculated once and for all using the DFPT capabilities of a code
package such as ABINIT.  All of the derived tensor properties
described in the following subsections can then be calculated from
these using simple matrix manipulations, as we shall see.

%-----------------------------------------------------------
\subsection{Relaxed-ion tensors}
\label{sec:relaxed}
%-----------------------------------------------------------

Generally, the physical static response properties of interest must
take into account the relaxations of the ionic coordinates. This
becomes especially important for non-centrosymmetric polar systems,
such as ferroelectric ones, where these various effects become
coupled. Instead of ``clamped-ion'' quantities $\chibar$, $\Cbar$ and
$\ebar$ defined at fixed $u$, we can define the corresponding
``relaxed-ion'' or ``dressed'' response tensors $C$, $\chi$, and
$e$ as follows. To develop expressions for these, we let
\begin{equation}
\widetilde{E}(\eta,\E)=\min_u E(u,\E,\eta)
\;.
\label{eq:Emin}
\end{equation}
Referring back to Eq.~(\ref{eq:Eexpan}), setting $\partial E/\partial u_n=0$,
$\partial E/\partial\E_\alpha=0$, and $\partial E/\partial \eta_j=0$,
and assuming that the reference configuration is
one in which the forces $A_m$ vanish, we find
\[ 0=B_{nm}\,u_m+B_{n\alpha}\,\E_\alpha+B_{nj}\,\eta_j \]
from which it follows that
\begin{equation}
u_m=-(B^{-1})_{mn} \,[\,B_{nj}\,\eta_j+B_{n\alpha}\,\E_\alpha]
\;.
\end{equation}
Defining
\begin{eqnarray}
\chi_{\alpha\beta} = - \, \frac{\partial^2\widetilde{E}}
   {\partial\E_\alpha \partial\E_\beta} \, \Bigg\vert_{\eta}
\; ,
\label{eq:chidef}
\\
C_{jk} = \frac{\partial^2\widetilde{E}}
   {\partial\eta_j \partial\eta_k}\, \Bigg\vert_{\E}
\; ,
\label{eq:Cdef}
\\
e_{\alpha j} = - \, \frac{\partial^2\widetilde{E}}{\partial\E_\alpha \partial\eta_j}
\; ,
\label{eq:edef}
\end{eqnarray}
and using Eqs.~(\ref{eq:Kdef}-\ref{eq:ebardef}), we find that the
physical relaxed-ion dielectric, elastic, and piezoelectric tensors
become
\begin{eqnarray}
\chi_{\alpha\beta} &=& \chibar_{\alpha\beta}
   + \Omega_0^{-1}\,Z_{m\alpha}\,(K^{-1})_{mn}\,Z_{n\beta}
\; ,
\label{eq:chi}
\\
C_{jk} &=& \Cbar_{jk}-\Omega_0^{-1}\,\Lambda_{mj}\,
(K^{-1})_{mn}\,\Lambda_{nk}
\; ,
\label{eq:C}
\\
e_{\alpha j} &=& \ebar_{j\alpha}+\Omega_0^{-1}\,
Z_{m\alpha}\,(K^{-1})_{mn}\,\Lambda_{nj}
\; ,
\label{eq:e}
\end{eqnarray}
respectively.

Note that Eqs.~(\ref{eq:chi}-\ref{eq:e}) cannot be naively
evaluated as written because the force-constant matrix $K$ is
singular, due to the fact that $K$ has three vanishing eigenvalues
associated with translational symmetry.  Moreover, in soft-mode
systems, other eigenvalues may be close to zero, and care should
be taken to distinguish these from the translational ones.
For these reasons, we have implemented a
careful procedure for obtaining the ``pseudo-inverse'' of $K$;
throughout these notes, whenever we refer to $K^{-1}$, we really
mean the pseudo-inverse.

We proceed as follows. (i) We identify the three-dimensional space
of acoustic modes (i.e., uniform translations), and construct a
$(3N)\times(3N)$ orthogonal matrix $U$ whose first
three columns correspond to these translational modes; the remaining
columns of $U$ are formed by Graham-Schmidt orthogonalization of the
basis. (ii) We construct $K'=UKU^T$, whose upper $3\times3$
block represents the acoustic subspace and therefore ought to be
zero.
(iii) We let $K'_{\rm red}$ be the lower $(3N-3)\times(3N-3)$
block of $K'$, corresponding to the reduction to the complementary
subspace of optical modes. (iv) We invert $K'_{\rm red}$ by
standard means (taking appropriate measures in case
this matrix is nearly singular, as when soft modes have nearly
vanishing frequencies).
Let the result be
denoted as $(K^{-1})'_{\rm red}$. (v) We pad $(K^{-1})'_{\rm red}$
with zeros in the first three rows and columns to form the
$(3N)\times(3N)$ matrix $(K^{-1})'$.  (vi) Finally, we define
the pseudo-inverse of $K$ to be $K^{-1}=U^T(K^{-1})'U$.

Thus, by construction, the resulting pseudo-inverse $K^{-1}$
is zero in the subspace of translational modes, and is the inverse
of the original matrix in the complementary subspace.
As a result, any time $K^{-1}$ is multiplied by another tensor,
a pre-projection onto the complementary subspace of dimension $3N-3$
is effectively carried out.  In other words, the acoustic
sum rule is effectively enforced in any operation involving
$K^{-1}$.

%-----------------------------------------------------------
\subsection{Other derived tensor quantities}
\label{sec:other}
%-----------------------------------------------------------

In the previous subsection, we showed how to obtain the static
dielectric susceptibility tensor $\chi_{\alpha\beta}$,
the elastic tensor $C_{jk}$, and the piezoelectric
tensor $e_{\alpha j}$.  These quantities are defined
under conditions of controlled strain and electric field.  From
these three ingredients, it is straightforward to form many other
useful tensor quantities describing physical properties defined
under other constraints or boundary conditions, as we shall see in
this section.

%-----
\subsubsection{Dielectric tensors}
%-----

The susceptibility tensor $\chi_{\alpha\beta}$ is defined
at fixed (vanishing) strain; the corresponding dielectric tensor is
\begin{equation}
\eps^{(\eta)}_{\alpha\beta}=\epsilon_0\left(\delta_{\alpha\beta}+
    \chi_{\alpha\beta}\right)
\;,
\label{eq:chieta}
\end{equation}
where $\epsilon_0$ is the susceptibility of free space (SI units are used
throughout) and the superscript $(\eta)$ indicates that the derivative is taken
at fixed strain.  Often one is interested instead in the free-stress
dielectric tensor
\begin{equation}
\eps^{(\sigma)}_{\alpha\beta}=\epsilon_0\left(\delta_{\alpha\beta}+
    \chi^{(\sigma)}_{\alpha\beta}\right)
\label{eq:epsdef}
\end{equation}
which incorporates the free-stress susceptibility
$\chi^{(\sigma)}$.  An expression for the latter is
easily derived from the elastic enthalpy
\begin{equation}
\widetilde{H}(\sigma,\E)=\min_{\{\eta\}} \, \big[
 \, \widetilde{E}(\eta,\E) - \eta_j \sigma_j \, \big]
\end{equation}
Following a line of reasoning similar to that leading from
Eq.~(\ref{eq:Emin}) to Eqs.~(\ref{eq:chi}-\ref{eq:e})
and setting $\sigma_j=0$, one obtains
\begin{equation}
\chi^{(\sigma)}_{\alpha\beta}=\chi_{\alpha\beta}
 +e_{\alpha j}\,(C^{-1})_{jk}\,e_{\beta k}
\; .
\label{eq:chifs}
\end{equation}
Typically, an AC dielectric measurement will access the true static
susceptibility $\chi^{(\sigma)}$ as long as the frequency
is much less than that of sample resonances (elongational, bending,
or torsional modes), and $\chi^{(\eta)}$ at frequencies much
higher than sample resonances (but much less than phonon frequencies).

Before leaving this subsection, we note that it is convenient to
define inverse dielectric tensors
\begin{eqnarray}
\beta^{(\eta)} &=& (\eps^{(\eta)})^{-1} \;,
\label{eq:betaeta}
\\
\beta^{(\sigma)} &=& (\eps^{(\sigma)})^{-1}
\label{eq:betasig}
\end{eqnarray}
for later use.

%-----
\subsubsection{Elastic and compliance tensors}
%-----

The elastic tensor $C_{jk}$ defined in Sec.~\ref{sec:relaxed}
is the one defined under conditions of fixed (vanishing) electric field:
$C_{jk}=C_{jk}^{(\E)}$.  It may sometimes
be of interest to treat instead the elastic tensor
$C_{jk}^{(D)}$ defined under conditions of fixed
electric displacement field ${\bf D}$.  For example, in the case
of a thin film of dielectric material sandwiched between much thicker
layers of other insulating host materials, the electrostatic boundary
conditions fix the component of $D$, not $\E$, normal to the
interfaces.  One readily obtains
\begin{equation}
C^{(D)}_{jk}=C^{(\E)}_{jk}
+e_{\alpha j}\,
\beta^{(\eta)}_{\alpha\beta} \,e_{\beta k}
\; .
\label{eq:ela_D}
\end{equation}

It is also straightforward to obtain the corresponding elastic
compliance tensors either under zero $\E$-field
\begin{equation}
S^{(\E)}=(C^{(\E)})^{-1}
\label{eq:Ecompliance}
\end{equation}
or under zero $D$-field
\begin{equation}
S^{(D)}=(C^{(D)})^{-1}
\label{eq:Dcompliance}
\end{equation}
boundary conditions.

%-----
\subsubsection{Piezoelectric tensors}
\label{sec:piezo}
%-----

The formulation of an energy functional appropriate to the
simultaneous treatment of strains and electric fields is rather
subtle, as discussed in the Appendix.  There, we show that the
proper (relaxed-ion) piezoelectric tensor $e_{jk}$ introduced in
Sec.~\ref{sec:relaxed} may be written as
\begin{equation}
e_{\alpha j} ={\partial P_\alpha\over\partial\eta_j} \Big\vert_\E
\label{eq:eone}
\end{equation}
or equivalently, by a thermodynamic relation,\cite{Nye,Ballato}
\begin{equation}
e_{\alpha j} =-{\partial \sigma_j \over\partial\E_\alpha} \Big\vert_\eta
\; ,
\label{eq:etwo}
\end{equation}
where it is understood (see Appendix) that in these and subsequent
equations, $\bf P$ and $\E$ are to be interpreted as the reduced
polarization ${\bf P}'$ and electric field $\E'$ and of
Eqs.~(\ref{eq:Pprime}) and (\ref{eq:Eprime}), respectively.
This is done so that Eqs.~(\ref{eq:eone}) and (\ref{eq:etwo}) will
yield the proper, rather than the improper, piezoelectric tensor.\cite{proper}

In view of  Eq.~(\ref{eq:etwo}), $e_{\alpha j}$ is sometimes referred
to as a ``piezoelectric stress constant.''  In any case, it is the natural
piezoelectric constant defined under conditions of controlled $\E$
and $\eta$.
On the other hand, the ``piezoelectric strain constant''
$d_{\alpha j}$, defined under conditions of controlled $\E$ and $\sigma$,
is equally or even more commonly discussed in the literature; it is
defined via
\begin{equation}
d_{\alpha j}= {\partial\eta_j\over\partial\E_\alpha} \Big\vert_\sigma
\end{equation}
or
\begin{equation}
d_{\alpha j} = {\partial P_\alpha\over\partial\sigma_j} \Big\vert_\E
\end{equation}
and is given in terms of $e$ via
\begin{equation}
d_{\alpha j}=S^{(\E)}_{jk}\,e_{\alpha k}
\label{eq:pie_d}
\end{equation}
as can again be derived from ther\-mo\-dy\-na\-mic re\-la\-tions.
\cite{Nye,Ballato}
Two other, less commonly used, piezoelectric tensors are $g_{\alpha j}$
and $h_{\alpha j}$, defined under conditions of fixed $(D,\sigma)$ and
$(D,\eta)$, respectively, and given by
\begin{eqnarray}
g_{\alpha j}&=&\beta^{(\sigma)}_{\alpha\beta}\,d_{\beta j} \;,
\label{eq:pie_g}
\\
h_{\alpha j}&=&\beta^{(\eta)}_{\alpha\beta}\,e_{\beta j} \;,
\label{eq:pie_h}
\end{eqnarray}
where the $\beta$ are the inverse dielectric tensors defined in
Eqs.~(\ref{eq:betaeta}-\ref{eq:betasig}).  These have the properties
\begin{eqnarray}
\delta \eta _j &=& g_{\alpha j} \,\delta D_\alpha \;,  \nonumber\\
\delta \E_\alpha  &=& -g_{\alpha j} \, \delta \sigma _j \;, \nonumber\\
\delta \sigma _j &=&-h_{\alpha j} \, \delta D_\alpha  \;, \nonumber\\
\delta \E_\alpha  &=& -h_{\alpha j} \,\delta \eta _j  \;.
\end{eqnarray}
%

%-----
\subsubsection{Piezoelectric coupling coefficients}
\label{sec:coupfac}
%-----

The most common definition of the piezoelectric coupling factor
$k_{\alpha j}$is given by\cite{Ballato,Standard}
\begin{equation}
k_{\alpha j}=  \frac{\vert d_{\alpha j}\vert}{\sqrt{
  \eps_{\alpha\alpha}^{(\sigma)}
  \, S_{jj}^{(\E)} }}
\;.
\label{eq:kfac}
\end{equation}
This applies to the case where the field is applied only along $\alpha$
and the only non-zero stress is the one with Voigt label $j$.  For
example, $k_{33}$ is a dimensionless measure of the coupling of
electric and strain degrees of freedom along the $\hat z$ axis.
Roughly speaking, a coupling factor close to unity implies an excellent
impedance match for the material used as an electromechanical
transducer between the specified electric and elastic channels
(a coupling factor greater than unity is forbidden by stability
considerations).\cite{Ballato}

Note that $k_{\alpha j}$ in Eq.~(\ref{eq:kfac}) does not transform
like a tensor, and the usual implied sum notation does not apply to
this equation.  Instead, we can define a tensorially correct,
dimensionless coupling tensor via
\begin{equation}
{\cal K}=[\,\beta^{(\sigma)}]^{1/2} \cdot d \cdot [C^{(\E)}]^{1/2}
\;,
\label{eq:kappa}
\end{equation}
where an obvious matrix-product notation is used.  The standard
``singular-value decomposition'' can be used to write $\cal K$ as
\begin{equation}
{\cal K}={\cal U} \cdot
\begin{pmatrix}
   \tilde{k}_1 & 0 & 0 & 0 & 0 & 0 \cr
   0 & \tilde{k}_2 & 0 & 0 & 0 & 0 \cr
   0 & 0 & \tilde{k}_3 & 0 & 0 & 0 \cr
\end{pmatrix}
\cdot {\cal V}^T
\;,
\label{eq:svd}
\end{equation}
defined by requiring that $\cal{U}$ and $\cal{V}$ be orthogonal
3$\times$3 and 6$\times$6 matrices respectively, and that the
$\tilde{k}_\nu$ should be positive.
(Alternatively, the $\tilde{k}_\nu^2$ can be determined as the eigenvalues
of the 3$\times$3 symmetric matrix ${\cal K\,K}^T =
[\,\beta^{(\sigma)}]^{1/2} \, d \,
C^{(\E)} \, d^T \, [\,\beta^{(\sigma)}]^{1/2}$.)  For each
singular value, the corresponding columns of $\cal U$ and $\cal V$
give the pattern of electric field and of strain, respectively,
that are directly coupled to one another by $\cal K$.

The coupling factors can be related to differences between dielectric
or compliance tensors defined under different boundary conditions.
Starting from Eqs.~(\ref{eq:epsdef}-\ref{eq:chifs}) and
Eqs.~(\ref{eq:ela_D}-\ref{eq:Dcompliance}), one can show
\begin{eqnarray}
\epsilon^{(\sigma)} - \epsilon^{(\eta)}
   &=& d\,C^{({\cal E})}  d^T
     = [\epsilon^{(\sigma)}]^{1/2} \,{\cal K} \, {\cal K}^T
         \, [\epsilon^{(\sigma)}]^{1/2}
\;,
\label{eq:diffeps}
\\
S^{({\cal E})} - S^{(D)}
   &=& d^T  \beta^{(\sigma)}   d
     = [S^{({\cal E})}]^{1/2} \, {\cal K}^T {\cal K}
         \, [S^{({\cal E})}]^{1/2}
\;.
\label{eq:diffS}
\end{eqnarray}
Specializing to high-symmetry situations in which $\epsilon$ is
necessarily diagonal, one finds, for example,
\begin{equation}
\frac{\epsilon^{(\sigma)}_{\alpha\alpha}
-\epsilon^{(\eta)}_{\alpha\alpha}}
{\epsilon^{(\sigma)}_{\alpha\alpha}} = \tilde{k}_\alpha^2
\;.
\label{eq:ediff}
\end{equation}
%

%%%%%%%%%%%%%%%%%%%%%%%%%%%%%%%%%%%%%%%%%%%%%%%%%%%%%%%%%%%%%%%%%%%%%%%
\section{Methods and details of the calculations}
\label{sec:methods}
%%%%%%%%%%%%%%%%%%%%%%%%%%%%%%%%%%%%%%%%%%%%%%%%%%%%%%%%%%%%%%%%%%%%%%%

Our {\it ab-initio} calculations were carried out using the ABINIT code
package.\cite{ABINIT,explan-avail}
Specifically, we first carried out full structural
relaxations for both materials. Next, response-function calculations
were carried out in order to obtain first derivatives of the occupied
wavefunctions with respect to the perturbations of atomic displacements
(i.e., phonons at $q=0$), uniform electric field, and strain.
These were then used to compute the elementary second-derivative
response-function tensors, Eqs.~(\ref{eq:Kdef}-\ref{eq:ebardef}),
of Sec.~\ref{sec:elementary}.  Except for the diagonal elements of some
elementary tensors, this was done using a non-variational
expression that only requires input of one of the two corresponding
wavefunction derivatives.\cite{Gonze}
(As for the ``mixed derivative''
tensors $\Lambda$, $e$ and $Z$, strain derivatives were used for
$\Lambda$ and $e$, and displacement derivatives were used for
$Z$.\cite{explan-dudk}) Finally, from
these elementary response tensors, the various secondary response
tensors of Secs.~\ref{sec:relaxed} and \ref{sec:other} are obtained
according to the formulas given there.  All calculations are at
zero temperature.

The DFT and DFPT calculations for ZnO and BaTiO$_3$ were carried
out using Troullier-Martins pseudopotentials\cite{Troullier}
and a plane-wave energy cutoff of 50 hartree.  
The Zn pseudopotential includes
the $3d$ electrons in the valence, as this has been shown to be
important for accurate results.\cite{Ref1} An $8\times 8\times 8$
and $6\times 6\times 6$ Brillouin-zone k-point sampling were used
for ZnO and BaTiO$_3$ respectively. The exchange and correlation
effects were treated within the local-density approximation (LDA) 
in the Ceperley-Alder\cite{Ceperley} form with the
Perdew-Wang\cite{Perdew} parameterization.

Finally, we made one additional modification in the case of BaTiO$_3$,
where it is well known that the usual underestimation of the
equilibrium lattice constant associated with the local-density
approximation has an unusually profound
influence on the ferroelectric distortion, which is very sensitive
to cell volume.\cite{Zhong}  Therefore, to get more physically
meaningful results, we carried out the initial structural relaxation
with the cell volume constrained to be that of the experimental
structure at zero temperature.  This is similar in spirit to the
use of a ``negative fictitious pressure'' that is a standard
feature of many first-principles based studies of ferroelectric
perovskite materials.\cite{Zhong}

%%%%%%%%%%%%%%%%%%%%%%%%%%%%%%%%%%%%%%%%%%%%%%%%%%%%%%%%%%%%%%%%%%%%%%%
\section{Results for two sample systems: Z\lowercase{n}O and
B\lowercase{a}T\lowercase{i}O$_3$}
%%%%%%%%%%%%%%%%%%%%%%%%%%%%%%%%%%%%%%%%%%%%%%%%%%%%%%%%%%%%%%%%%%%%%%%

In this section, we consider two paradigmatic systems, wurtzite ZnO
and rhombohedral BaTiO$_3$. The electromechanical properties of ZnO
make it a widely used material in mechanical actuators and piezoelectric
sensor applications, while BaTiO$_3$ is a prototypical perovskite
ferroelectric material.  It is of particular interest to compare and
contrast the behavior of these two materials in view of the fact that
BaTiO$_3$ is a soft-mode system, while ZnO is not.  This may help
provide insight into the role of the soft mode, which can be expected
to lead to enhanced piezoelectric and dielectric response and enhanced
couplings.  We first describe the results of our ground-state DFT
calculations, and then present the results for the various linear-response
tensors as defined in Sec.~\ref{sec:form}.

Because ZnO is not a soft-mode system, its properties depend only
weakly on temperature, so that it is not unreasonable to compare
room-temperature experimental results with zero-temperature theory.
BaTiO$_3$ is a different matter, as its properties depend
crucially on temperature.  The room-temperature tetragonal
phase has indeed been thoroughly studied, and as a result,
its dielectric, elastic, and piezoelectric constants have been
systematically measured.\cite{Berlincourt,Zgonik}  However,
there are formidable difficulties associated with preparing
single-crystal, single-domain samples of the low-temperature
rhombohedral phase, and of carrying out dielectric and elastic
measurements on such samples at low temperature under well-defined
electric and elastic boundary conditions.  As a result,
almost no reliable experimental values are available for the
corresponding materials constants at very low temperature.
Therefore, for the purposes of this paper, we have adopted
the approach of providing comparisons with experiment for ZnO
wherever possible to benchmark our approach, and of presenting
our calculations of the low-temperature properties of BaTiO$_3$
as predictions for a system that is difficult to characterize
experimentally.

%-----------------------------------------------------------
\subsection{Relaxed structural properties}
%-----------------------------------------------------------

%-----
\subsubsection{ZnO}
%-----

The ground state of ZnO is a tetrahedrally coordinated wurtzite
structure (space group $P6_3mc$, point group $C_{6v}$) with
four atoms per unit cell. The structure is determined by three
parameters: the hexagonal edge $a$, the height of the prism $c$,
and the internal parameter $u$.  The structural results from our
full relaxation are given in Table \ref{tab:ZnO_stru}.
For comparison, an ideal wurtzite with exactly tetrahedral angles
and equal-length bonds would have $u=3/8$ and $c/a=\sqrt{8/3}$.
As is typical of DFT calculations, we find that the lattice constants
are underestimated by 1-2\%.

\begin{table}[t]
\caption{Structural parameters of ZnO.  Lattice constants $a$ and
$c$ in \AA; $u$ is dimensionless internal parameter.}
\label{tab:ZnO_stru}
\begin{tabular}{ldddd}
     & $a$ & $c$ & $c/a$ & $u$ \cr
\hline
Present work & 3.197 & 5.166 & 1.616 & 0.380 \cr
Previous theory\tablenotemark[1] & 3.286 & 5.241  & 1.595 & 0.383 \cr
Expt.\tablenotemark[2]  & 3.250 & 5.210  & 1.602 & 0.382 \cr
\end{tabular}
\tablenotetext[1] {Ref.~\onlinecite{Catti}.}
\tablenotetext[2] {Ref.~\onlinecite{ZnO-exp}.}
\end{table}

\begin{table}[b]
\caption{Relaxed structure of rhombohedral BaTiO$_3$.  Lattice
constant $a$ and atomic displacements $\Delta$ (relative to ideal
cubic positions) in \AA; rhombohedral angle $\theta$ in degrees.}
\begin{tabular}{lrrrrr}
\label{tab: BTO_stru}
& $a$ & $\theta$ & $\Delta_z$(Ti) & $\Delta_x$(O) & $\Delta_z$(O)  \cr
\hline
 Theory & 4.00 & 89.85 & 0.043 & $-$0.049 & $-$0.077  \cr
 Expt.\tablenotemark[1] & 4.00 & 89.90 & 0.052$\pm$12 &
$-$0.044$\pm$8 & $-$0.072$\pm$8 \cr
\end{tabular}
\tablenotetext[1] {Ref.~\onlinecite{BTO-exp}.}
\end{table}

%-----
\subsubsection{BaTiO$_3$}
%-----

BaTiO$_3$ is a prototypical example of the class of perovskite
ferroelectric materials.  These materials normally have the
paraelectric cubic perovskite structure at high temperature, but
then undergo a ferroelectric instability as the temperature is
reduced.  BaTiO$_3$ actually goes through a series of three
ferroelectric phase transitions as the symmetry is first
tetragonal, then orthorhombic, and then rhombohedral, with
polarization respectively along [100], [110], and [111], with
decreasing temperature.  The ground-state rhombohedral structure
(space group $R3m$, point group $C_{3v}$) is fully
determined by its lattice constant, rhombohedral angle, and
the symmetry-allowed internal atomic displacements
along the $[111]$ direction. We represent
the rhombohedral phase in the hexagonal coordinate system,
in which the $z$ axis is along the previous $[111]$ direction.
Table \ref{tab: BTO_stru} lists the structural parameters of
our relaxed BaTiO$_3$, in which we constrained the atomic
volume to be equal to the experimental one as explained
in Sec.~\ref{sec:methods}.  The remaining structural parameters
can be seen to be in good agreement with experiment.

%-----------------------------------------------------------
\subsection{Displacement response tensors}
%-----------------------------------------------------------

%-----
\subsubsection{ZnO}
%-----

Wurtzite ZnO belongs to space group $P6_3mc$ ($C^{4}_{6v}$). Standard
group-theory analysis shows that the $\Gamma$-point phonon modes can be
decomposed as
\begin{equation}
  \Gamma_{\rm opt}=A_1 \oplus 2B_1 \oplus E_1 \oplus 2E_2
\end{equation}
in which the $A_1$ and $E_1$ modes are both Raman and IR active,
while the nonpolar $E_2$ modes are Raman active only and $B_1$
modes are silent. Shown in Table \ref{tab:ZnO-phonon} are our
computed phonon frequencies compared with two experimental
results, showing good agreement with experiment.

\begin{table}[t]
\caption{Frequencies(in cm$^{-1}$) of the zone-center optical
phonon modes in wurtzite ZnO.}
\label{tab:ZnO-phonon}
\begin{center}
\begin{tabular}{lddd}
  symmetry character  & Theory &
  Expt.\cite{ZnO-phonon1}  & Expt\cite{ZnO-phonon2} \cr
\hline
 A$_1$(TO) &390 &380 & 380 \cr
 A$_1$(LO) &548 & 574 & 579 \cr
 B$_1$     &261 & --- & --- \cr
 E$_1$(TO) &409 & 407 & 413 \cr
 E$_1$(LO) &552 & 583 & 591 \cr
 E$_2$  &  91 & 101 & 101 \cr
 E$_2$  &  440 & 437 &444 \cr
\end{tabular}
\end{center}
\end{table}

Because of the wurtzite symmetry of ZnO, the effective charge tensor
$Z$ has only two independent elements, while the force-response
internal-strain tensor $\Lambda$ has only four independent elements.
We present results for both tensors in Table \ref{tab:ZnO_displacement}. For this
semiconductor material, it can be seen that the effective charge is
very close to the nominal ionic charge.

\begin{table}[t]
\caption{Independent elements of the Born effective charge tensor $Z$
(in units of $e$) and of the force-response internal strain tensor
(hartree/bohr) for wurtzite ZnO.}
\label{tab:ZnO_displacement}
\begin{center}
\begin{tabular}{lddddd}
     $Z_{xz}({\rm Zn})$ & $Z_{zz}({\rm Zn})$ \cr
2.135 & 2.163 \cr
\hline
$\Lambda_{x5}({\rm Zn})$ & $\Lambda_{x6}({\rm Zn})$ &
$\Lambda_{z3}({\rm Zn})$ & $\Lambda_{y1}({\rm O})$ \cr
     $-$9.5 & $-$15.0 & 18.7 & $-$16.7 \cr
\end{tabular}
\end{center}
\end{table}

%-----
\subsubsection{BaTiO$_3$}
%-----

The low-temperature phase of BaTiO$_3$ has a rhombohedral structure
which belongs to the $R3m$ space group. According to a group-theory
analysis, the zone-center phonon frequencies can be decomposed as
\begin{equation}
  \Gamma_{\rm opt}=3{\rm A}_1 \oplus 4{\rm E} \oplus {\rm A}_2 \;.
\end{equation}
The A$_1$ and E modes are both IR and Raman active, while the A$_2$
mode is silent.  Table \ref{tab:BTO-phonon} gives the calculated
phonon frequencies at the $\Gamma$ point.
(The A$_2$ mode at 278\,cm$^{-1}$ and the E modes at 293\,cm$^{-1}$
are the ones derived from the silent F$_{2u}$ modes of the undistorted
cubic structure; because the distortion is weak, the LO--TO splitting
of these E modes is negligible.)
The results are very similar to those of the previous theoretical
study of Ghosez.\cite{Ghosez-thesis}
While we are not aware of detailed experimental
information about phonon frequencies at very low temperature, we note
that measurements just below the orthorhombic to rhombohedral phase
transition temperature indicate phonon frequencies in
three regions (100-300 cm$^{-1}$, 480-580 cm$^{-1}$, and 680-750 cm$^{-1}$)
in qualitative agreement with our zero-temperature calculations.

\begin{table}[b]
\caption{Phonon frequencies (in cm$^{-1}$) at the $\Gamma$ point for
rhombohedral BaTiO$_3$.}
\label{tab:BTO-phonon}
\begin{tabular}{ldld}
Phonon mode & Frequency & Phonon mode & Frequency \cr
\hline
A$_1$(TO1) & 169 & E(TO1) & 164 \cr
A$_1$(LO1) & 179 & E(LO1) & 175 \cr
A$_1$(TO2) & 255 & E(TO2) & 206 \cr
A$_1$(LO2) & 460 & E(LO2) & 443 \cr
A$_1$(TO3) & 511 & E(TO3) & 472 \cr
A$_1$(LO3) & 677 & E(LO3) & 687 \cr
A$_2$      & 278 & E      & 293 \cr
\end{tabular}
\end{table}

We also calculated the atomic Born effective charges for this phase,
but in view of the lower symmetry and larger number of independent
elements, we have not listed them all here (our
results are again very similar to those of
Ref.~\onlinecite{Ghosez-thesis}).  The cation results are easily
given as
$Z_{xx}$(Ba) = $Z_{yy}$(Ba) = 2.78, $Z_{zz}$(Ba) = 2.74,
$Z_{xx}$(Ti) = $Z_{yy}$(Ti) = 6.64, and $Z_{zz}$(Ti) = 5.83.
The effective charge tensors are non-diagonal and non-symmetric for
the oxygens; we mention only that the
eigenvalues of the symmetric parts of these tensors cluster around
$-$2 and $-$5, i.e., not much changed from their cubic-phase values.
Similarly, we have computed the full internal-strain tensor
$\Lambda$ for rhombohedral BaTiO$_3$, but we have chosen not to
present the details here because of the complicated form of this
tensor involving a large number of independent elements.

%-----------------------------------------------------------
\subsection{Dielectric tensors}
\label{sec:diel}
%-----------------------------------------------------------

%
\begin{table}[t]
\caption{Dielectric tensors of ZnO and BaTiO$_3$ (in units
of $\epsilon_0$).}
\label{tab:diel}
\begin{center}
\begin{tabular}{lcddddd}
& & \multicolumn{3}{c}{Present theory} & \multicolumn{2}{c}{Experiment} \cr
& Index & $\bar{\epsilon}$ & $\epsilon$ & $\epsilon^{(\sigma)}$
& $\bar{\epsilon}$ & $\epsilon$ \cr
\hline
ZnO       & 11 & 5.76 & 10.31 &  11.09  & 3.70\tablenotemark[1] &
     7.77\tablenotemark[1] \cr
          & 33 & 5.12 & 10.27 &  12.67  & 3.78\tablenotemark[1] &
     8.91\tablenotemark[1] \cr
BaTiO$_3$ & 11 & 6.20 & 68.75 &  264.75 & 6.19\tablenotemark[2] & ---\cr
          & 33 & 5.79 & 37.44 &  49.51  & 5.88\tablenotemark[2] & ---\cr
\end{tabular}
\end{center}
\tablenotetext[1] {Ref.~\protect\onlinecite{Ashkenov}.}
\tablenotetext[2] {Ref.~\protect\onlinecite{Wang}.}
\end{table}

We now turn to a discussion of the computed dielectric tensors
for wurtzite ZnO and rhombohedral BaTiO$_3$, which are presented
in Table \ref{tab:diel}.
Because of the high point-group symmetry, the dielectric tensors
have only two independent elements.  Recall that the clamped-ion tensor
$\bar\epsilon$, the fixed-strain relaxed-ion tensor $\epsilon$,
and the free-stress relaxed-ion tensor $\epsilon^{(\sigma)}$ are defined
through Eqs.~(\ref{eq:chi}) and (\ref{eq:chieta}-\ref{eq:chifs}).
While the results for the purely electronic dielectric tensors
$\bar{\epsilon}$ are in good agreement with experiment for BaTiO$_3$,
we find that our LDA theory significantly overestimates the electronic
dielectric response of ZnO.  Hill and Waghmare,\cite{Ref1} also
using an LDA pseudopotential approach, found $\bar{\epsilon}_{33}=4.39$,
not as large as our 5.76, but still much larger than the experimental
3.70.  At least some of this overestimate is undoubtedly attributable
to the LDA (and is connected with the underestimate of the gap in LDA),
but the choice of pseudopotentials also seems to play a role.
The computed lattice contributions
$\epsilon_{11}-\bar{\epsilon}_{11}=4.55$ and
$\epsilon_{33}-\bar{\epsilon}_{33}=5.15$ are in better agreement with
the experimental values of 4.07 and 5.13 respectively.

While the clamped-ion tensors
$\bar{\epsilon}$ are not so different for these two materials,
the lattice contribution is clearly much bigger for the BaTiO$_3$ case.
That is, while the lattice contribution $(\epsilon-\bar\epsilon)$
is about the same size as the purely electronic one $(\bar\epsilon)$
for ZnO, it is almost 10 times larger in BaTiO$_3$.  This difference
clearly reflects the fact there is a soft ferroelectric mode present
in the latter material.  (Here, we use ``soft'' in the sense of a
mode that has a small, but positive frequency; it is, of course,
closely related to the imaginary-frequency unstable mode computed
for the cubic structure, which condenses out to form the ferroelectric
rhombohedral phase.)  In the semiconductor ZnO, on the other hand,
no such soft mode is present.

The last column presents our results for the free-stress dielectric
tensors $\epsilon^{(\sigma)}$ that are related to the fixed-strain
tensors $\epsilon$ via Eq.~(\ref{eq:chifs}).
The tensors $\epsilon$ and $\epsilon^{(\sigma)}$ are
the same in higher-symmetry crystals, but in the presence of piezoelectric
coupling, they are, in general, different.  The free-stress tensors
are always larger than the fixed-strain ones because the additional
strain relaxation occurs so as to allow further polarization to
develop in the direction of the applied field.  We can see that the
changes are modest for ZnO (on the order of 10-20\%), which is not
a soft-mode system.  On the other hand, they are much more
profound for the case of BaTiO$_3$, where most notably an
order-of-magnitude change occurs for $\epsilon_{11}$.  This is
related to the large value of the piezoelectric coupling factor
$k_{15}$, as we will see later in Sec.~\ref{sec:coupfac}.
Essentially, it arises because the polarization vector is rather
soft with respect to rotation away from the $z$ axis, so that the
electric susceptibility is large in the $x$-$y$ plane.

%-----------------------------------------------------------
\subsection{Elastic tensors}
%-----------------------------------------------------------

We now consider the various elastic tensors.  Recall that
the clamped-ion elastic tensor $\bar C$ of Eq.~(\ref{eq:Cbardef}) is just
the second derivative of the unit-cell energy with respect
to homogeneous strains, without allowing for internal structural
relaxations, while the physical elastic tensor $C$ of
Eq.~(\ref{eq:C}) does include such relaxations.
This tensor $C$ (written more explicitly as $C^{({\cal E})}$) usually
defined under conditions of fixed macroscopic
electric field, but it is sometimes of interest to consider the
elastic tensor $C^{(D)}$ of Eq.~(\ref{eq:ela_D}) defined instead
under conditions of fixed electric displacement field.
These are identical for higher-symmetry (e.g., centrosymmetric) crystals,
but that is not the case here.
The compliance tensors $S$ are defined as the inverses of the
corresponding elastic tensors $C$.

\begin{table*}[t]
\caption{Clamped-ion ($\bar{C}$) and relaxed-ion
($C$) elastic tensors at constant $\cal E$, relaxed
ion ($C^{(D)}$) elastic tensor at constant $D$ (GPa), and
corresponding compliance tensors ($\bar{S}$, $S$, and
$S^{(D)}$) (TPa$^{-1}$), for wurtzite ZnO. Previous
theoretical and experimental results are also given for $C$ for
comparison.}
\label{tab:ela_ZnO}
\begin{center}
\begin{tabular}{lddddddcc}
& \multicolumn{6}{c}{Present theory} &
Theo.\tablenotemark[1] & Expt.\tablenotemark[2] \cr
Index & $\Cbar$ & $C$ & $C^{(D)}$ & $\Sbar$ & $S$ & $S^{(D)}$ &
$C$ & $C$ \cr
\hline
11 & 305 & 226 & 231 & 3.86 & 7.79 & 7.56          & 246 & 209 \cr
12 & 107 & 139 & 144 & $-$1.20 & $-$3.63 & $-$3.93 & 127 & 120 \cr
13 & 77 & 123 &  114 & $-$0.61 & $-$2.12 & $-$1.58 & 105 & 104 \cr
33 & 333 & 242 & 260 & 3.29 & 6.28 & 5.23          & 246 & 211 \cr
44 & 62 & 40 & 43 & 16.23 & 24.69 & 23.21          &  56 &  44\cr
66 & 99 & 44 & 44 & 10.12 & 22.84 & 22.73          & --- &  --- \cr
\end{tabular}
\end{center}
\tablenotetext[1] {Ref.~\protect\onlinecite{Catti}.}
\tablenotetext[2] {Ref.~\protect\onlinecite{Landolt}.}
\end{table*}
\begin{table}
\caption{Clamped-ion ($\bar{C}$) and relaxed-ion
($C$) elastic tensors at constant $\cal E$, relaxed
ion ($C^{(D)}$) elastic tensor at constant $D$ (GPa), and
corresponding compliance tensors ($\bar{S}$, $S$, and
$S^{(D)}$) (TPa$^{-1}$), for rhombohedral BaTiO$_3$.}
\label{tab:ela_BTO}
\begin{center}
\begin{tabular}{ldddddd}
Index & $\Cbar$ & $C$ & $C^{(D)}$
      & $\Sbar$ & $S$ & $S^{(D)}$ \cr
\hline
11 & 349 & 277 & 318 & 3.32 & 5.85 & 3.65 \cr
12 & 106 & 79 & 93 & $-$0.82 & $-$2.94 & $-$0.95 \cr
13 & 96 & 41 & 81 & $-$0.72 & $-$0.45 & $-$0.68 \cr
14 & 8.4 & 45 & 19 & $-$0.31 & $-$8.17 & $-$0.89 \cr
33 & 334 & 264 & 323 & 3.41 & 3.93 & 3.44 \cr
44 & 110 & 48 & 97 & 9.12 & 35.85 & 10.63 \cr
65 & 8.3 & 45 & 19 & $-$0.63 & $-$16.33 & $-$1.78 \cr
66 & 121 & 99 & 113 & 8.28 & 17.58 & 9.18 \cr
\end{tabular}
\end{center}
\end{table}

The results of our calculations of these tensors are displayed in
Table \ref{tab:ela_ZnO} and \ref{tab:ela_BTO} for ZnO and BaTiO$_3$
respectively.  The lower point-group symmetry of BaTiO$_3$ is reflected
in the presence of an additional symmetry-allowed element $C_{14}$.
Actually, there are only five independent elements for ZnO, since
$C_{66}=(C_{11}-C_{12})/2$ and $S_{66}=2(S_{11}-S_{12})$
by symmetry.\cite{Nye}  Similarly, there are really only six independent
elements for BaTiO$_3$; in addition to the same relation, one
also has
$C_{66}=(C_{11}-C_{12})/2$, $S_{66}=2(S_{11}-S_{12})$,
$C_{56}=C_{14}$, and $S_{56}=2C_{14}$.
Our results for the elastic constants of ZnO can be seen to be in good
agreement with previous theory and with experiment (last columns
of Table \ref{tab:ela_ZnO}).

We notice that the physical elastic $C_{jk}$ are generally
smaller than the frozen-ion ones ${\bar C}_{jk}$ (at least
for diagonal elements), since the additional internal relaxation
allows some of the stress to be relieved.  By the same token,
diagonal $S$ values are larger than $\bar S$ ones, reflecting the
increased compliance allowed by the relaxation of the atomic coordinates.
As for the dielectric constants, the differences are substantially
smaller for ZnO than for BaTiO$_3$, as a result of the soft-mode
contribution in the latter material.
The constraint of fixed electric displacement field has the effect
of suppressing some of this internal relaxation (for essentially the same
reason that longitudinal optical phonons are stiffer than
transverse optical ones).  This additional stiffness results
in larger diagonal $C^{(D)}$ values than $C$ values, and lower
diagonal $S^{(D)}$ values than $S$ values.  However, the differences
between $C^{(D)}$ and $C$ tensors are generally smaller than
the differences between $C$ and $\bar C$ tensors, especially for
ZnO.

%-----------------------------------------------------------
\subsection{Piezoelectric tensors}
%-----------------------------------------------------------

The bare (or ``frozen-ion'') piezoelectric tensor $\bar{e}_{\alpha
j}$ is just given by the mixed second derivatives of unit cell
energy with respect to electric field ${\cal E}_\alpha$ and strain
$\eta_j$, deforming internal atomic coordinates in strict
proportion to the homogeneous strain.  The full
piezoelectric tensor $e_{\alpha j}$ also takes into account the
contributions from the lattice, as described in Eq.~\ref{eq:e}. The
total number of independent piezoelectric tensor members is
determined by the point group of material. Rhombohedral BaTiO$_3$
(point group C$_{3v}$) has a lower symmetry than that of wurtzite
ZnO (C$_{6v}$), so we may expect more independent elements in the
former.  Indeed, a symmetry analysis\cite{Nye} shows that
ZnO has only three independent tensor elements, namely $e_{31}$
and $e_{33}$ describing polarization
along the $c$ axis induced by uniaxial $c$-axis or biaxial
$ab$-plane strains, while $e_{24}$ describes the polarization
induced by shear strains. For BaTiO$_3$, the symmetry is slightly
lower, and as a result there is a fourth independent tensor element
in this case.

In Table \ref{tab:piezo_ZnO} and \ref{tab:piezo_BTO}, we present
our results for piezoelectric tensors for these two materials.  We
also also give the results for the $d_{\alpha j}$ tensor as defined
in Eq.~\ref{eq:pie_d}. Our results for the $e_{\alpha j}$ matrix
for ZnO are consistent with the previous theory.\cite{Ref1,Ref2,Ref3}
(Table \ref{tab:piezo_BTO} shows five tensor elements for BaTiO$_3$,
not four, but in fact $e_{21}=e_{16}$ and $2d_{21}=d_{16}$ by
symmetry.\cite{Nye})
\begin{table*}[t]
\caption{Clamped-ion $\bar{e}$ (C/m$^2$), relaxed-ion $e$ (C/m$^2$),
and relaxed-ion $d$ (pC/N) piezoelectric tensors for
wurtzite ZnO. Previous theoretical and experimental 
results are also given for $e$ and $d$ for comparison.}
\label{tab:piezo_ZnO}
\begin{center}
\begin{tabular}{ldddcccc}
& \multicolumn{3}{c}{Present theory} &
\multicolumn{2}{c}{Theo.\tablenotemark[1]} & 
\multicolumn{2}{c}{Expt.\tablenotemark[2]} \cr
Index & $\ebar$ & $e$ & $d$ & $e$ & $d$ & $e$ & $d$ \cr
\hline
31 & 0.37 & $-$0.67 & $-$5.5 & $-$0.55 & $-$3.7 & $-$0.62 & $-$5.1  \cr
33 & $-$0.75 & 1.28 & 10.9 & 1.19 & 8.0 & 0.96 & 12.3  \cr
15 & 0.39 & $-$0.53 & $-$13.1 & $-$0.46 & $-$8.2 & $-$0.37 & $-$8.3  \cr
\end{tabular}
\end{center}
\tablenotetext[1] {Ref.~\protect\onlinecite{Catti}.}
\tablenotetext[2] {Ref.~\protect\onlinecite{Landolt}.}
\end{table*}

Recall that the frozen-ion and relaxed-ion piezoelectric tensors
are defined by Eqs.~(\ref{eq:ebardef}) and (\ref{eq:e}), in which the
relaxed-ion piezoelectric tensor incorporates contributions from
lattice relaxation. For the same reason as discussed previously
for the case of the dielectric and elastic tensors, the difference
between the above two tensors (e.g. $\bar{e}$
and $e$) is much bigger for BaTiO$_3$ than for ZnO, as expected
from the presence of the soft mode in the perovskite material.
Also, note that the electronic and lattice contributions have
opposite signs, with the lattice contribution being the larger
of the two, as is common for other dielectric materials.

In view of this partial cancellation of terms of opposite sign,
accurate calculations of $e$ and $d$ coefficients are particularly
delicate.  We find that our results for the 31 and 33 elements of the
$e$ and $d$ coefficients of ZnO are in reasonably good
agreement with experiment (slightly better than previous
Hartree-Fock calculations \cite{Catti}),
whereas we somewhat overestimate the
shear coefficients $e_{15}$ and $d_{15}$ (slightly more so than in
the Hartree-Fock theory\cite{Catti}).

\begin{table}
\caption{Clamped-ion $\bar{e}$ (C/m$^2$), relaxed-ion $e$ (C/m$^2$),
and relaxed-ion $d$ (pC/N) piezoelectric tensors for
rhombohedral BaTiO$_3$.}
\label{tab:piezo_BTO}
\begin{tabular}{lddd}
Index & $\ebar$ & $e$ & $d$ \cr
\hline
21 & $-$0.23 & 2.91 & 70.1 \cr
31 & 0.05 & $-$3.03 & $-$6.8 \cr
33 & $-$0.19 & $-$4.44 &$-$14.7 \cr
15 & 0.18 & $-$5.45 & $-$243.2 \cr
16 & $-$0.23 & 2.91 & 140.2 \cr
\end{tabular}
\end{table}
%

%-----------------------------------------------------------
\subsection{Electromechanical coupling constants}
%-----------------------------------------------------------

We compute and present in Table \ref{tab:kfactors} the
piezoelectric coupling factors $k_{33}$, $k_{31}$, and $k_{15}$
defined in Eq.~(\ref{eq:kfac}) for both ZnO and BaTiO$_3$.
We also calculate the singular values $\tilde{k}_\nu$ of the $\cal K$
matrix of Eq.~(\ref{eq:kappa}).  Because of the axial symmetry,
these are arranged into a pair of degenerate values $\tilde{k}_1=\tilde{k}_2$
corresponding to in-plane electric fields, and a $\tilde{k}_3$ corresponding
to axial fields.  (In fact, due to the symmetry, $\tilde{k}_\nu$ cab be
calculated in practice as just
$[\beta^{(\sigma)}_{\nu\nu}\,(d\cdot C^{(\E)} \cdot d^T)_{\nu\nu}]^{1/2}$.)
These $\tilde{k}_\nu$ values are also given in Table \ref{tab:kfactors}.

\begin{table}
\caption{Dimensionless piezoelectric coupling factors.
The first three correspond to $\E$-fields longitudinal to
the crystal axis; the last two are transverse.  Coupling
constants $\tilde{k}_1$ and $\tilde{k}_3$ are obtained from singular-value
analysis of the coupling tensor $\cal K$ (see text).}
\label{tab:kfactors}
\begin{center}
\begin{tabular}{lddd}
& ZnO & BaTiO$_3$ \cr
\hline
$k_{33}$& 0.41 & 0.35 \cr
$k_{31}$& 0.19 & 0.13 \cr
$\tilde{k}_3$& 0.44 & 0.49 \cr
$k_{15}$& 0.27 & 0.84 \cr
$\tilde{k}_1$& 0.27 & 0.86 \cr
\end{tabular}
\end{center}
\end{table}

Roughly speaking, the couplings given in the first three lines of
Table \ref{tab:kfactors} are associated with symmetry-preserving
``polarization stretching'' degrees of freedom, while those in
the last two lines correspond to ``polarization rotation'' modes.
Note that $k_{15}=\tilde{k}_1$ for ZnO but not for BaTiO$_3$;
this is a feature of symmetry, arising because an electric field
$\E_1$ couples uniquely to $\eta_5$ in ZnO, but also to $\eta_6$
in BaTiO$_3$.  Also, we can see that $\tilde{k}_1 \ge k_{15}$
and $\tilde{k}_4 \ge \max (k_{33},k_{31})$ in both materials,
since $\tilde{k}$ describes the optimal coupling between electric
and elastic channels.

Comparing the two materials, we see that the coupling factors
are rather comparable in the polarization stretching channel;
evidently, the soft mode does not play such a profound role there.
In contrast, the coupling factor $k_{15}$ is very large in
BaTiO$_3$; in fact, it is not far from unity, the maximum value
consistent with stability.  Indeed, this is precisely because
the crystal is not far from being unstable with respect to a
rotation of the polarization away from the rhombohedral axis
-- precisely the type of distortion that would carry it to the
orthorhombic phase, from which it evolved as the temperature was
reduced below the orthorhombic-rhombohedral phase transition
temperature that occurs experimentally at $\sim180\,$K.  The
large $k_{15}$ is also strongly connected to the large difference
between $\epsilon=\epsilon^{(\eta)}$ and $\epsilon^{(\sigma)}$
in Tab.~\ref{tab:diel} as already discussed at the end of
Sec.~\ref{sec:diel}.

The calculated value $k_{15}=0.84$ for BaTiO$_3$ is
quite respectable; it is in the range of the values of
$k_{15}=0.25$--$0.80$\cite{Cao} for the PMN-PT and PZN-PT  
single-crystal piezoelectrics on the rhombohedral 
side of the morphotropic phase boundary.
Unfortunately, the fact that this large coupling occurs only at
very low temperature probably makes it useless for practical
applications.  On the other hand, the present work suggests
that if a material like BaTiO$_3$ could somehow be stabilized in
the rhombohedral phase at room temperature, it might have very
promising piezoelectric properties.

%%%%%%%%%%%%%%%%%%
\section{SUMMARY}
%%%%%%%%%%%%%%%%%%%%%%%%%%%%%%%%%%%%%%%%%%%%%%%%%%%%%%%%%%%%%%%%%%%%%%%

In summary, we have developed a method that systematically treats
the effects of perturbations associated with atomic displacements,
electric fields and strains in insulating crystals, so that
physical quantities expressible as second derivatives of the total
energy can be computed efficiently.  In the first step, six elementary
tensors are computed once and for all using the methods of
DFPT: the force-constant matrix, the Born charge tensor, the
internal-strain tensor, and the frozen-ion dielectric, elastic,
and piezoelectric tensors.  The internal-displacement degrees
of freedom are then eliminated to give physical low-frequency
dielectric, elastic, and piezoelectric tensors, defined under
boundary conditions of controlled electric field $\E$ and strain
$\eta$.  We have also shown how these can then be manipulated to
obtain various related tensor properties of interest such
as the free-stress dielectric tensor, the fixed-$D$ elastic and
compliance tensors, and various piezoelectric tensors and
electromechanical coupling factors.  Such a systematic approach
is especially important in polar crystals, in which the
atomic-displacement, electric-field, and strain degrees of
freedom are strongly coupled in complex ways.

We have applied our approach to compute these tensor properties
for two paradigmatic crystals, ZnO and BaTiO$_3$, at zero temperature.
These materials differ most significantly in that there is a ferroelectric
soft mode that has condensed, but still remains rather low in frequency,
in the latter material.  The calculations are subject to several
approximations, most notably the LDA itself (and its associated
lattice-constant error, which has been removed by hand for the
case of BaTiO$_3$ -- see Sec.~\ref{sec:methods}), but also the frozen-core
approximation (as implemented through the use of pseudopotentials)
and the neglect of zero-point fluctuations.  Nevertheless,
we validate the approach by finding reasonably good agreement
between theory and experiment for most quantities in the case of ZnO,
despite the fact that the experiments are room-temperature ones.  The
largest discrepancies are for the purely electronic dielectric tensor
elements $\bar{\epsilon}_{11}$ and $\bar{\epsilon}_{33}$, the shear
piezoelectric coupling $e_{15}$, and to a lesser extent, derived
quantities that depend on these elementary ones.  In the
case of BaTiO$_3$, where low-temperature experiments on single-crystal,
single-domain samples under well-defined boundary conditions are
not available, our calculations provide useful predictions of the
material constants.  In particular, we find an encouraging
value of 0.84 for the $k_{15}$ electromechanical coupling constant,
and argue that this is associated with the proximity of the
orthorhombic phase.

We wish to emphasize that the usefulness of the general
approach advocated here transcends the particular implementation
of it (here DFT/LDA, pseudopotentials, etc.).  For example,
similar calculations might by carried out with Hartree-Fock methods
using localized orbitals\cite{Catti} or, eventually, using
``LDA+U'', dynamical mean-field theory, or quantum Monte Carlo methods.
In this case, the six elementary tensors of
Eqs.~(\ref{eq:Kdef}-\ref{eq:ebardef}) will first need to be
calculated using methods appropriate to the particular type of
electronic-structure method used.  However, the subsequent manipulations
described in Secs.~\ref{sec:relaxed} and \ref{sec:other} can then
be carried through in identically the same way as done here.

Finally, we note that the approach described here can be extended
to include other types of perturbations, such as alchemical ones,
and to the treatment of higher-order responses (e.g., anharmonic
elastic constants, nonlinear dielectric responses,
and electrostriction effects), providing possible directions
for future developments of the method.

\acknowledgments

The work was supported by ONR Grant N00014-97-1-0048
and by the Center for Piezoelectrics by Design under ONR
Grant N00014-01-1-0365.  We wish to thank K.~Rabe for useful
discussion, and M.~Veithen for a careful reading of the manuscript.

% ------------------------------------------
\appendix
\section{Simultaneous treatment of strains and electric fields}
\label{sec:app}
% ------------------------------------------

The formulation of an energy functional, and the definition of response
functions in terms of its second derivatives, is somewhat subtle in the
case that electric fields and strains are simultaneously present.  The
purpose of this Appendix it to give a careful treatment of the theory
in this case.  Except where noted, the notation here follows that
of Sec.~\ref{sec:relaxed} in that the internal displacements $u_m$
are assumed to have been integrated out (i.e., internal displacements
$u_m$ do not appear explicitly).

We begin introducing the deformation tensor
$\widetilde{\eta}_{\alpha\beta}$ in the Cartesian frame via
\begin{equation}
dr_\alpha=\widetilde{\eta}_{\alpha\beta}\,r_\beta
\label{eq:epstilde}
\end{equation}
(implied sum notation) where $dr_\alpha$ is the deformation of the
medium from its undeformed position $r_\alpha$.
We consider deformations
taking the form of homogeneous strains and rigid rotations, so that
the antisymmetry part of the deformation tensor
$\widetilde{\eta}_{\alpha\beta}$ describes the rotational part,
while its symmetric part is just the strain tensor
\begin{equation}
\eta_{\alpha\beta}=\frac{1}{2}\left(\widetilde{\eta}_{\alpha\beta}+
                               \widetilde{\eta}_{\beta\alpha}\right)
\;.
\label{eq:straindef}
\end{equation}
The {\it improper} piezoelectric tensor is defined as
\begin{equation}
e^{\rm impr}_{\alpha\beta\gamma}=
\frac{dP_\alpha}{d\widetilde{\eta}_{\beta\gamma}}
\;.
\label{eq:improper}
\end{equation}
The name ``improper'' reflects that fact that $e^{\rm impr}$
contains contributions that are spurious in a certain sense.\cite{proper}
For example, if we consider
a pure rotation of a spontaneously polarized crystal about an axis
that does not coincide with $\bf P$, then $\bf dP$ does
not vanish, and consequently $e^{\rm impr}_{\alpha\beta\gamma}$
has a component that is antisymmetric under interchange of
$\beta$ and $\gamma$.
Similarly, if we consider a uniaxial or biaxial compression in the
$x-y$ plane of a ferroelectric modeled as a lattice of discrete rigid
dipoles oriented in the $z$ direction, the polarization will change
even if the dipoles do not, because the polarization is defined as
the dipole moment per unit volume.  This, too, is an essentially
spurious effect.  By contrast, the ``proper'' piezoelectric tensor
will be defined so as to vanish in either of these situations.

In the presence of a strain, it is convenient to introduce {\it reduced}
or {\it rescaled} polarizations $P'_\alpha$ and electric fields
$\E'_\alpha$ via
\begin{equation}
P'_\alpha=(\widetilde{\eta}^{\;-1})_{\alpha\beta}\,P_\beta
\label{eq:Pprime}
\end{equation}
and
\begin{equation}
\E'_\alpha=\widetilde{\eta}_{\beta\alpha}\,\E_\beta
\;,
\label{eq:Eprime}
\end{equation}
where $P'_\alpha$ and $\E'_\alpha$ coincide with $P_\alpha$ and
$\E_\alpha$ in the absence of strains or rotations.
We then take our fundamental energy functional to be
\begin{equation}
\widetilde{E}(\E',\widetilde{\eta}) =\frac{1}{\Omega_0}
\left[ E^{(0)}_{\rm cell}-\Omega\,\E'\cdot{\bf P}'\right]
\label{eq:Etot}
\end{equation}
where $E^{(0)}_{\rm cell}$ is the usual zero-field Kohn-Sham
energy per cell\cite{explan-enthalpy} of the occupied Bloch functions
introduced earlier in Sec.~\ref{sec:elementary}.

Note that ${\bf P}\cdot\E={\bf P}'\cdot\E'$, so that Eq.~(\ref{eq:Etot})
is closely related to Eq.~(\ref{eq:Etotimprop}).  However,
it is important to understand that $\E'$ and $\widetilde{\eta}$
are the ``natural variables'' of the energy functional (\ref{eq:Etot}),
so that subsequent partial derivatives are defined in terms of this pair
of variables.  For example, the {\it proper} piezoelectric tensor is
now given by
\begin{equation}
e_{\alpha\beta\gamma}=
-\,\frac{\partial^2\widetilde{E}}{\partial\E'_\alpha\,
\partial\widetilde{\eta}_{\beta\gamma}}
\label{eq:properE}
\end{equation}
or
\begin{equation}
e_{\alpha\beta\gamma}=\frac{1}{\Omega_0}\,
\frac{\partial(\Omega P'_\alpha)}{\partial\widetilde{\eta}_{\beta\gamma}}
\;.
\label{eq:proper}
\end{equation}

We also emphasize that $\E'$ is, in many ways, a more natural variable
than $\E$ from the experimental point of view.  For example, if one
controlls the voltage $V$ across a film of $M$ atomic layers between
conducting capacitor plates and observes the resulting strain, one
is actually controlling $\E'=eV/Mc_0=\E c/c_0$, not $\E$, where $c_0$ and $c$
are the zero-field and finite-field lattice constants, respectively,
in the normal direction.

From Eqs.~(\ref{eq:improper}) and (\ref{eq:proper}) it follows that
the improper and proper piezoelectric tensors are related by\cite{proper}
\begin{equation}
e_{\alpha\beta\gamma}=
e^{\rm impr}_{\alpha\beta\gamma}-\delta_{\beta\gamma}\,P_\alpha
+\delta_{\alpha\beta}\,P_\gamma
\;.
\label{eq:compare}
\end{equation}
It is then easy to show that the proper tensor $e_{\alpha\beta\gamma}$
is symmetric under interchange of indices $\beta$ and $\gamma$, so that
the Voigt notation can be restored.  This is to be expected
because the reduced quantity ${\bf P}'$ is invariant under a
rigid rotation of the crystal, a fact that follows trivially from
Eq.~(\ref{eq:Pprime}).  That is, ${\bf P}'$, expressed as a function
of the six symmetrized strain variables (Eq.~(\ref{eq:straindef}))
and the three rotational variables, is actually independent of the
rotational ones.  Indeed, a rigid rotation of the entire
system, material plus external field $\E$, leaves both $\E'$ and
${\bf P}'$ individually unchanged.  It is then natural to discard
the rotational variables and recast the symmetric strain variables
in Voigt notation.  We then regard the energy functional of
Eq.~(\ref{eq:Etot}) to be a functional $\widetilde{E}(\E',\eta)$
of the fundamental variables of rescaled $\E'_\alpha$ and Voigt
$\eta_j$, and the proper piezoelectric tensor may be written as
\begin{equation}
e_{\alpha j}= -\,\frac{\partial^2\widetilde{E}}{\partial\E'_\alpha\,
\partial\eta_j} =\frac{1}{\Omega_0}\,\frac{\partial (\Omega P'_\alpha)}
{\partial\eta_j}
\;.
\label{eq:propertwo}
\end{equation}

Restoring the explicit dependence on internal displacements $u_m$,
Eq.~(\ref{eq:Etot}) becomes
\begin{equation}
E(u,\E',\eta) =\frac{1}{\Omega_0}
\left[ E^{(0)}_{\rm cell}-\Omega\,\E'\cdot{\bf P}'\right]
\;,
\label{eq:Etotfull}
\end{equation}
where $\widetilde{E}(\E',\eta)$ of Eq.~(\ref{eq:Etot}) corresponds to the
minimum of (\ref{eq:Etotfull}) over all possible displacements $u_m$.
While this equation is numerically equal to Eq.~(\ref{eq:Etotimprop}),
it is critical to recall that it is written in terms of
different arguments.

Strictly speaking, this notation should have been introduced at the
very beginning of Sec.~\ref{sec:elementary}, and every equation
throughout the paper, starting with Eq.~(1), should have $\E$
replaced by $\E'$ and $\bf P$ by ${\bf P}'$.  For example,
Eqs.~(\ref{eq:ebardef}) and Eq.~(\ref{eq:eone}) should be replaced by
Eqs.~(\ref{eq:properfull}) and
\begin{equation}
\bar{e}_{\alpha j}= -\,\frac{\partial^2 E}{\partial\E'_\alpha\,\partial\eta_j}
\;,
\label{eq:properfull}
\end{equation}
respectively, and
similarly for all other equations.  However, for the purposes
of clarity of presentation, it was decided to avoid use of this
clumsy notation in the main part of the paper.

Finally, we note that the reduced quantities
${\bf P}'$ and $\E'$ are also rather natural physical variables
from the point of view of computational implementation.  Indeed,
these two quantities can further be expressed in terms of {\it
fully reduced} quantities $p_\mu$ and $\varepsilon_\mu$ via
\begin{equation}
{\bf P}'=\frac{e}{\Omega}\,p_\mu\,{\bf a}_\mu^{(0)}
\label{eq:pred}
\end{equation}
and
\begin{equation}
\varepsilon_\mu=e\,\E'\cdot{\bf a}_\mu^{(0)}
\label{eq:ered}
\end{equation}
so that ${\bf P}=(e/\Omega)\,p_\mu{\bf a}_\mu$ and
$\varepsilon_\mu=e\,\E\cdot{\bf a}_\mu$.  In these equations,
${\bf a}_\mu^{(0)}$ is the $\mu$'th undeformed primitive real-space
lattice vector, ${\bf a}_\mu$ is the corresponding deformed lattice
vector, and $\Omega={\bf a}_1\cdot{\bf a}_2\times{\bf a}_3$ is the
deformed cell volume.  Note that ${\bf P}\cdot\E={\bf P}'\cdot\E'
= \Omega^{-1} \, p_\mu\varepsilon_\mu$ and
\begin{equation}
e_{\alpha j}=
\frac{e}{\Omega_0}\,a_{\mu,\alpha}\,\frac{\partial p_\mu}{\partial\eta_j}
\;.
\label{eq:piezofr}
\end{equation}
The fully reduced polarization
$p_\mu$ has a simple interpretation in terms of the fractional
positions of the charges in the unit cell; for example, the contribution
to it coming from filled band is just $-1/2\pi$ times the Berry
phase of that band, as can be seen by comparing with Eq.~(10)
of Ref.~\onlinecite{proper}. Similarly, $\varepsilon_\mu$ is just
$e$ times the electrostatic potential drop across the unit cell in
direction ${\bf a}_\mu$.

The computational implementation of DFPT is done
quite naturally in terms of these reduced quantities,
\cite{ABINIT,Hamann,explan-avail} and as a result, DFPT automatically
yields the proper piezoelectric tensor.\cite{Hamann} This can be
a source of confusion when comparing the DFPT results
with those of finite-difference calculations.  In the latter approach,
the polarization is obtained directly from ground-state DFT
calculations,\cite{ksv} and piezoelectric tensor elements are obtained
by numerical differentiation using sufficiently small strains about the
reference structure.  This procedure yields the improper tensor, however,
and Eq.~(\ref{eq:compare}) must be applied to compare such results
with the DFPT ones.\cite{Hamann}

%%%%%%%%%%%%%%%%%%%%%%%%%%%%%%%%%%%%%%%%%%%%%%%%%%%%%%%%%%

%%%%%%%%%%%%%%%%%%%%%%%%%%%%%%%%%%%%%%%%%%%%%%%%%%%%%%%%%%%


\begin{thebibliography}{0}
\bibitem{explan-enthalpy}
That is, the electrostatic part of $E^{(0)}_{\rm cell}$ is defined
to be $(\epsilon_0/2)\int_{\rm cell}d{\bf r}\,\E_{\rm loc}^2({\bf r})$,
where the local field $\E_{\rm loc}({\bf r})$ is the difference between
the microscopic field $\E({\bf r})$ and its unit-cell average $\E$.
\bibitem{explan-dudk}
In the case of $Z$ and $e$, the wavefunction derivatives with respect
to wavevector are also needed (see Ref.~\protect\onlinecite{Gonze}).
\end{thebibliography}
\end{document}